\documentstyle[prl,aps]{revtex}
\begin{document}
\draft
\input{psfig.sty}
\twocolumn[\hsize\textwidth\columnwidth\hsize\csname@twocolumnfalse\endcsname
\title{Oscillations and defect turbulence in a shallow fluidized bed}
\author{D.K.Clark$^1$, L.S. Tsimring$^1$, and I.S.Aranson$^2$}
\address{$^1$Institute for Nonlinear Science, University of California, 
San Diego, La Jolla, CA 92093-0402\\ $^2$ Materials Science Division,
Argonne National Laboratory, Argonne, IL 60439}
\date{\today}
\maketitle

\begin{abstract}
We report an experimental study of the dynamics of an air-fluidized thin
granular layer. Near-onset behavior of this shallow fluidized bed was
described in the earlier paper (Tsimring et al, 1999). Above the
threshold of fluidization the system exhibits a Hopf bifurcation as the
layer starts to oscillate at a certain frequency due to a feedback
between the layer dilation and the airflow drag force. After application of
temporal band-pass filtering of this frequency we discovered the
spatio-temporal dynamics in the form of defect turbulence. This type of
dynamics is natural for spatio-temporal systems close to the threshold
of a Hopf bifurcation.  At high flow rates,
low-frequency short-wavelength structures appear in addition
to the long-wavelength excitations. A simple model describing the
instability and occurrence of oscillations in a shallow fluidized bed,
is proposed.  
\end{abstract}

\pacs{PACS: 46.10.+z, 47.54.+r, 47.35.+i}

\narrowtext
\vskip1pc]

The non-equilibrium dynamics of granular materials have been a subject
of growing attention among physicists in the last few
years\cite{gran,swinney,shinbrot}.  Granular materials reveal a host of
interesting phenomena (compaction, self-organized criticality, patterns,
convection, etc.) when are subjected to external driving. A 
typical way of driving the granular medium out of equilibrium is
to fluidize it with an external gas or liquid flow. Such ``fluidized
beds'' are widely used in the industry for mixing solid and liquid
chemicals\cite{davidson,squires}.  One of the biggest problems
associated with this use of fluidized beds is an instability which leads
to macroscopic inhomogeneities and fluctuations in the bulk (bubbling,
slugging). At large flow rates, this leads to a developed
three-dimensional turbulent regime resembling boiling liquid.

In the recent paper\cite{TRS}, we proposed to study the dynamics of
air-fluidized granular materials in a bed with a large aspect ratio
between the horizontal size and the thickness of the layer. In this
geometry, the dynamics become quasi-two-dimensional and can be
investigated by imaging the surface of the granular layer.  In
Ref.\cite{TRS}, we studied the onset of fluidization and found that it
has the features of a phase transition, with very long transient
fluctuations just below the fluidization threshold.  In this paper, we
report the results of the experimental study of the thin granular layer
dynamics at high airflow rates, above the threshold of fluidization.  We
find that above this threshold the layer begins to oscillate at a
certain frequency which depends on the layer thickness and the airflow
rate.  Processing of surface images revealed the occurrence of
propagating waves which form small disordered short-lived spirals. At
larger airflow rate, irregular small-scale oscillating cellular pattern
appears on the surface of the bed.  The temporal frequency of these
oscillating bubbles is significantly less than that of large-scale
waves. We believe that these small-scale perturbations are  generated by
the large amplitude, long-wave layer oscillations in a way similar to
the mechanism of Faraday instability in fluid or granular layers subject
to external vertical vibrations (see, e.,g.,\cite{swinney,gollub}).  The
mechanism of layer oscillations is related to the dependence of the drag
force from the air flow acting on the granular layer, on the volume
fraction of particles in the layer. When the layer is dilated, the drag
force diminishes, and the layer falls back due to gravity. This picture
leads us to the formulation of the  simple dynamical model which
qualitatively agrees with our experimental results.

The experimental setup is similar to the one used in Ref.\cite{TRS}, but
it features a larger round porous bronze plate (diameter 15 cm,
thickness 0.5 cm, average pore size 6$\mu$m).  As a fluidizing fluid  we
use compressed dry air.  The granular material consists of monodisperse
spherical bronze particles of size 0.15 mm, and the bed depth was varied
from 3 to 7 particles deep.  The patterns on the surface of the granular
layer were illuminated using the low-angle light from a circular
fluorescent light around the bed, and recorded using high-speed Kodak
SR-C digital CCD camera.  We also used a pressure transducer SenSym
SX01DP1 to record the airflow fluctuations near the free surface of the
granular layer.

When the flow rate is below certain threshold value, the layer remains
completely static. The value of this critical airflow depends
sensitively on the grain parameters (size, shape, density) as well as on
the thickness of the layer. Unless explicitly noted otherwise,  we will
be referring to the bed thickness 0.65 mm.  For that thickness, the
critical airflow speed was 12.1 cm/sec. Slightly above the threshold,
the fluidization of the layer is highly non-uniform. As described in our
earlier paper Ref.\cite{TRS}, the fluidization occurs in small localized
regions surrounded by completely static grains. 

\begin{figure}
\centerline{\psfig{figure=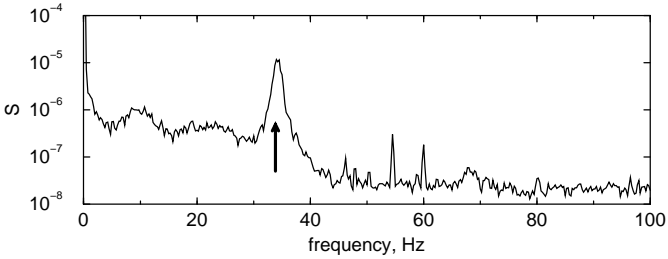,width=3.5in}}
\caption{Power spectrum of the pressure fluctuations near the layer
at the airflow rate 15.8 cm/sec. The arrow indicates the peak
corresponding to the large-scale layer oscillations.}
\label{fig1}
\end{figure}

\begin{figure}
\centerline{\psfig{figure=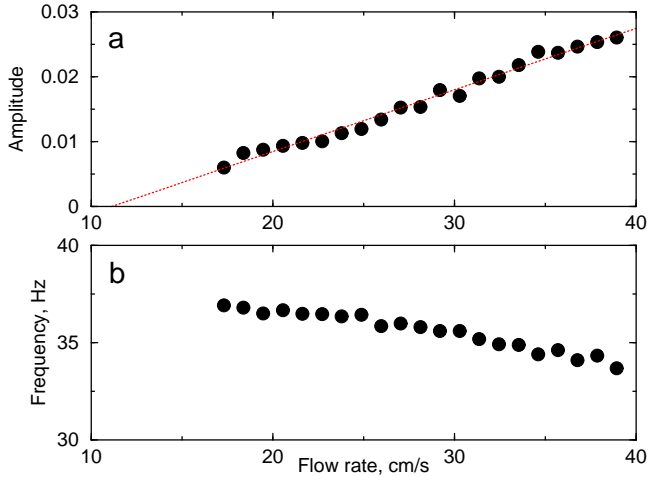,height=2.3in}}
\caption{Amplitude (a) and frequency (b) of the spectral peak in the
power spectrum of layer oscillations as a function of the
airflow rate $V$ for the layer thickness 0.65 mm. Linear regression of
the amplitude (dashed line in a) yields the oscillation threshold
of 11.2 cm/s, which is slightly lower than the fluidization threshold
12.1 cm/s}
\label{fig2}
\end{figure}

\begin{figure}
\centerline{\psfig{figure=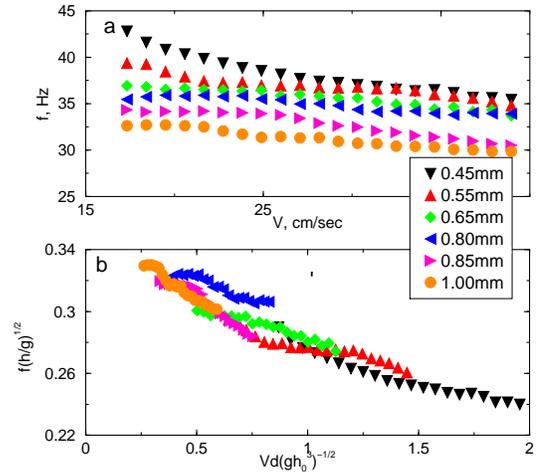,height=2.5in}}
\caption{a - frequency
$f$ of the spectral peak in the
power spectrum of layer oscillations as a function of the
airflow rate $V$; b - rescaled frequency $f(h/g)^{1/2}$ as a function of 
$Vd/(gh_0^3)^{1/2}$}
\label{fig3}
\end{figure}

At higher values of the airflow  the whole layer becomes fluidized and
exhibits noticable vertical oscillations.  The power spectrum of the pressure 
fluctuations
near the free surface of the layer is shown in Figure \ref{fig1}. We
measured the peak frequency and the magnitude as a function of the
airflow velocity (Figure \ref{fig2}). The amplitude of the oscillations
increases linearly and the frequency $f$ slightly decreases as
the airflow speed increases. Figure \ref{fig3}a shows dependence of the
peak frequency on the airflow speed for different values of layer
thickness. A naive scaling of variables by the gravity $g$ and the
thickness of the layer $h_0$ ($f(h_0/g)^{1/2}$, $V(gh_0)^{-1/2}$) does
not converge these data to a single curve.  It indicates that additional
non-dimensional parameters involving the particle size $d$ and the air
viscosity $\nu$ play an important role in the mechanism of these
oscillations. For example, if one chooses scaling $Vd/(h_0^3g)^{1/2}$ for
the airflow, then the data lie rather close to a single curve (see
Fig.\ref{fig3}b).


\begin{figure}
\centerline{\psfig{figure=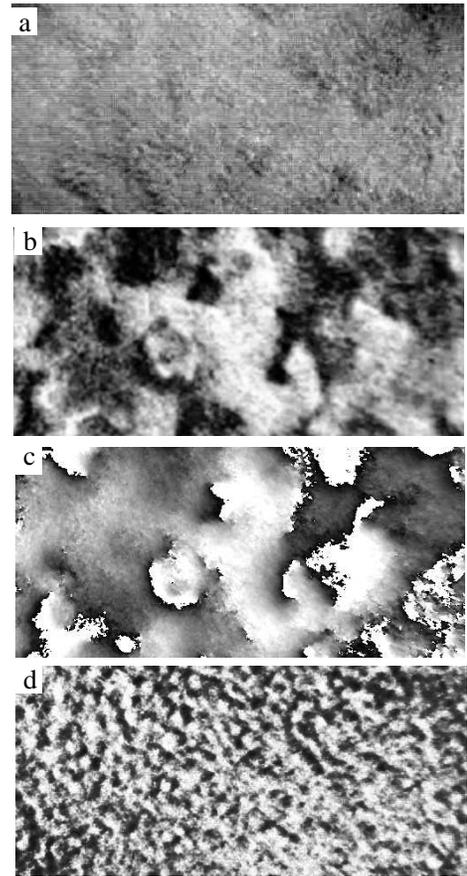,width=2.4in}}
\caption{Spatial structure of the 0.65 mm layer oscillations:
a - snapshot of the granular layer surface for $V=$15.8 cm/sec; 
b - corresponding snapshot from the temporal sequence filtered by 
the narrow-band filter with central frequency $f=35$ Hz; 
c - spatial distribution of the phase of 35 Hz oscillations;
d - snapshot from the temporal sequence for $V=$42.2 cm/sec filtered 
by a low-pass filter with the cutoff frequency $25$ Hz.}
\label{fig5}
\end{figure}

High-speed imaging of the surface  at 500 fps also confirms the
occurrence of the layer oscillations.  In order to determine the spatial
structure of these oscillations, we performed temporal filtering of the
recorded image sequences. We used a band-pass filter to extract
oscillations at the peak frequency $f$ for each pixel, and then combined
these filtered signals back to create the sequence of
temporally-filtered images\cite{swinney1}.

A raw snapshot from the video sequence taken at the airflow speed 15.8
cm/sec, and the corresponding filtered snapshot are shown in Figure
\ref{fig5}a,b.  Viewing filtered images in sequence\cite{mpeg} reveals
that the oscillations form propagating waves, moreover, these waves are
organized in small irregular spirals.  The sense of direction of the
wave propagation can be determined from the spatial distribution of the
phase of oscillations at the filtered frequency $f$ (Figure
\ref{fig5}c). Singularities of the phase correspond to the cores of the
spirals, or topological defects.  The phase distribution changes over
time, as the defects move, annihilate, and new defects are born.  Such
``defect turbulence'' is to be expected near the onset of the Hopf
bifurcation in spatially extended systems. It is generically described
by the Ginzburg-Landau equation for the complex amplitude of the
oscillation\cite{GL}. Similar defect turbulence has been observed in
spatially-extended Belousov-Zhabotinsky reaction\cite{BZ}.

\begin{figure}
\centerline{\psfig{figure=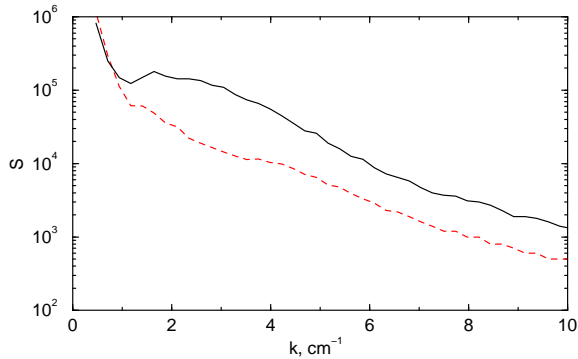,width=3.in}}
\caption{Azimuthally-averaged spatial power spectrum of oscillations at
the dominant frequency 31.2 Hz (dashed line) and low-frequency
oscillations $f<25$ Hz (solid line) for the airflow speed 42.2 cm/sec}
\label{fig8}
\end{figure}

At higher air flow, in addition to large-scale layer oscillations,
irregular small-scale perturbations of the layer appear.  These
perturbations correspond to the low-frequency part of the layer
oscillations spectrum. Figure \ref{fig5}d shows the snapshot of the
surface image at the airflow $42.2$ cm/sec obtained by low-pass
filtering of the image sequence with the cut-off frequency 25 Hz.  The
spatial power spectrum of these perturbations averaged over angle is
shown in Figure \ref{fig8} together with the corresponding spectrum for
the peak frequency 31 Hz. As one can see, spatial spectrum of
low-frequency oscillations exhibit a broad peak at wavenumber $k\approx
1.6$ cm$^{-1}$, whereas the spectrum of 31 Hz oscillations decays with
the wavenumber.

The mechanism of the shallow fluidized bed oscillations can be described
as follows. The layer dynamics are determined by the interplay of two
forces: gravity and the drag force from the gas flow. While gravity $g$
remains constant, the drag force varies depending on the volume fraction
of particles in the layer.  When the layer lies on the surface and is
densely packed, the drag force is maximal. On the other hand, when the
layer is lifted in the air and dilated, it becomes more ``transparent"
for the airflow, and the drag force decreases. This effect is known for
deep fluidized beds\cite{batchelor}. In the simplest form, we could
assume that the drag force $\Gamma$ is inversely proportional to the height
of the layer $h$, $\Gamma=\gamma/h$, where constant $\gamma$ depends on the
nominal flow speed $V$, close-packed thickness of the layer $h_0$, gas
viscosity, and the properties of the individual particles (size, shape).
Then the second Newton's law yields

\begin{equation}
h_{tt}=\frac{\gamma}{\rho h}-g,
\label{mod1}
\end{equation}
where subscript $t$ denotes time derivative, and $\rho$ is the total mass 
of grains per unit area. In non-dimensional variables $H=h/h_0$ and
$T=(g/h_0)^{1/2}t$, this equation  reads
\begin{equation}
H_{TT}=\frac{P}{H}-1,
\label{mod1n}
\end{equation}
where $P=\gamma/\rho g h_0$.  The amplitude and frequency of periodic
oscillations  for this nonlinear oscillator equation depend on the
initial conditions.  Assuming that the layer always comes down to the
surface where it has a minimal thickness $h_0$, we impose initial
conditions $H(0)=1$ and $H_T(0)=0$. With these initial conditions,
amplitude $A$ and frequency $\Omega=\omega(h_0/g)^{1/2}$  of
oscillations are uniquely determined by the parameter $P$. For $P$
slightly above 1, the amplitude of oscillations $A\approx P - 1$ and the
frequency is $\Omega\approx P^{-1/2}$. So, in qualitative agreement with
the experiment, near the threshold the amplitude increases approximately
linearly with $P$, and the frequency slowly decreases with $P$.  

While the model Eq.(\ref{mod1}) explains the nature of the layer
oscillations, it does not address the mechanism of the instability which
leads to the oscillations.  Indeed, since there is always some
dissipation due to friction and inelastic particle collisions, the
oscillations would decay, and the stable equilibrium bed thickness at
which the drag force balances gravity, would be reached.  To overcome
dissipation, an additional instability mechanism  is needed to excite
the layer oscillations.  The mechanism of this instability is related
to the known mechanism of instability for deep fluidized
beds\cite{batchelor}, in which it is caused by the inertia of particles in the
gas flow and leads to the time delay between the volume fraction variations and
the drag force changes.  Similarly, we can assume that the drag force
$\Gamma$ is not enslaved to the thickness the layer $h$, but approaches
$\gamma/ h$ asymptotically.  This additional degree of freedom can be
described by the equation $\alpha \Gamma_t=\gamma h^{-1} - \Gamma$, where
parameter $\alpha$ controls the rate of inertia in the dependence
between $\Gamma$ and $h$. We shall also modify the equation for the layer
thickness dynamics (\ref{mod1}) to include a thickness-dependent
dissipation. Indeed, the amount of dissipation varies strongly as a
function of the layer thickness, as it affects the frequency of
inelastic collision among grains. As the thickness approaches the
close-packed limit, the number of collisions, and the dissipation rate,
diverge. We assume that this effect is captured by an additional
dissipative term $-\sigma h_t/(h-h_0)^2$ (the actual power of
singularity at $h=h_0$ in not important here).  Now, the extended model in
non-dimensional form reads

\begin{eqnarray}
H_{TT}&=&F-1-\frac{\sigma' H_T}{(H-1)^2},
\label{mod2} \\
\alpha' F_T&=&\frac{P}{H} - F,
\label{mod3}
\end{eqnarray}
where $F=\Gamma/g\rho, \alpha'=\alpha(g/h_0)^{1/2}, \sigma'=\sigma(gh_0^3)^{-1/2}$.
Additional relaxation dynamics described by Eq.(\ref{mod3})
directly leads to an instability of the thin fluidized layer. 
Linearizing Eqs.(\ref{mod2}),(\ref{mod3}) near the fixed point $F=1,\ H=P$ 
yields the dispersion equation for perturbations $\propto \exp(-i\Omega T)$,
\begin{equation}
(1-i\alpha\Omega)\left[\Omega^2+\frac{i\sigma\Omega}{(P-1)^2}\right]=\frac{1}{P},
\label{disp}
\end{equation}
where we dropped primes at the non-dimensional parameters $\alpha$ and $\sigma$.
At small $\alpha$ and $\sigma$, the complex frequency of layer oscillations is
given by
\begin{equation}
\Omega\approx P^{-1/2}
+\frac{i}{2}\left[\frac{\alpha}{P}-\frac{\sigma}{(P-1)^2}\right].
\label{freq}
\end{equation}
As one can see from this formula, the instability always occurs at large
enough $P$.  Notice that for small $\alpha$ and $\sigma$, the solution
of Eqs.(\ref{mod2}),(\ref{mod3}) at large $T$ approaches that of
Eq.(\ref{mod1n}) with initial conditions $H(0)=1;\ H_T(0)=0$.  This is
not surprising, since these conditions are enforced by diverging
dissipation rate at $H=1$, and during the rest of the oscillation cycle,
non-conservative corrections are negligibly small. 

To conclude, in this paper we studied the dynamics of a shallow
fluidized granular layer.  Experiments with air-driven shallow fluidized
bed showed that above fluidization threshold, the layer exhibits 
oscillations at a certain frequency. These oscillations are in fact
propagating waves which have a structure of small disordered spirals. At
larger airflow speeds, in addition to the high-frequency spiral waves,
low-frequency short-wavelength perturbations are observed in the bed.
It is feasible that these small ``bubbles'' are formed via subharmonic
excitation by the primary high-frequency oscillations, as in the Faraday
instability.  We proposed a simple dynamical model describing
high-frequency oscillations in which the layer is driven by the airflow
drag force and gravity. The drag force in turn depends on the volume
fraction of particles in the layer, and therefore on the height of the
layer surface. The underlying assumptions for this model are based on
experimental facts and the intuitive physical picture, and a systematic
derivation of the model from the first principles may yield more
complicated relations among the layer parameters. Still, this model
describes the instability leading to the layer oscillations, and agrees
on a qualitative level with the observed dependence of the oscillation
magnitude and frequency on the airflow rate.  A more detailed study of
the model including its systematic derivation and generalization towards
spatiotemporal dynamics will be the subject of a separate publication.

Authors are grateful to A.Didwania,T.Shinbrot, and H.Swinney for useful
discussions.  This research was supported by the Energy Research Program
of the Office of Basic Energy Sciences at the US Department of Energy
(grants \# DE-FG03-95ER14516 and DE-FG03-96ER14592).

\end{document}